\documentstyle{article}\begin{document}
\title{Conformally flat travelling plane wave solutions of Einstein equations }
\author{ Z. Haba\\
Institute of Theoretical Physics, University of Wroclaw,\\ 50-204
Wroclaw, Plac Maxa Borna 9, Poland,\\
email:zbigniew.haba@uwr.edu.pl} \maketitle
\begin{abstract} We discuss conformally flat plane wave solutions
of Einstein equations depending on  the plane wave phase
$\xi=\omega\tau-{\bf qx}$, where $\tau$ is the conformal time. We
show that ideal fluid Einstein equations and scalar fields with
exponential self-interaction have solutions of this form. We
consider in more detail the source depending on $\xi$ with
$\omega=\vert{\bf q}\vert$ describing models of a massless scalar
field, electromagnetic field and relativistic particles with
space-time depending mass density. We obtain explicit conformally
flat metrics solving Einstein equations with such a source of the
energy-momentum .
\end{abstract}

\section{Introduction}
 On a large scale the
universe looks isotropic and homogeneous. Then, the dynamics
involves  only the expansion scale factor $a$. The resulting
$\Lambda$CDM model describes well \cite{LCDM} the observational
data..
 There is however some tension concerning the
value of the Hubble constant resulting from (local) CMB and
supernova observations. It may be that the problem can be
explained by inhomogeneities observed on a local scale
\cite{H}\cite{H2}\cite{ellis}. An assumption of the isotropy and
homogeneity allows to derive explicit solutions of Einstein
equations \cite{solutions}. Inhomogeneous cosmological models with
a spherical symmetry have been extensively studied (see the review
in \cite{bolejko}). At an intermediate scale there are some
phenomena (voids,  walls ) which disturb the homogeneous isotropic
picture \cite{cmb}\cite{ellis}. For a recent review of anisotropic
Bianchi cosmologies see \cite{kumar}. Particular solutions of
Einstein equations may have a cosmological meaning reflecting some
observed inhomogeneities and anisotropies in galaxy distribution
and in CMB. In this paper we find anisotropic and inhomogeneous
solutions of Einstein equations resulting from scalar and
electromagnetic fields as a source of plane waves. We make an
assumption that the metric is determined by the energy-momentum
evolving like a plane wave in a direction ${\bf q}$, i.e., that it
depends on $\xi=\omega\tau-{\bf qx}$, where $\tau $ is the
conformal time. We consider conformally flat metrics ( see
\cite{infeld} for their cosmological relevance). Then, its scale
factor $a$ also depends on $\xi$. We did not encounter  such an
explicit assumption in general relativity although the plane-wave
Ansatz is a standard tool in classical theory of scalar waves
(solitons)\cite{solitons}. Einstein equations  with a given lhs
could be treated as a definition of the energy-momentum on the
rhs. This is the way the Riemannian geometry is exploited in the
plasma physics \cite{fluids}. However, without a local Lagrangian
defining the energy-momentum on the rhs we would in general get
non-local and acausal theories. Plane-wave solutions of Einstein
equations are discussed (and classified) from the
group-theoretical point of view in \cite{solutions} (sec.37). We
think that such plane-waves, when their velocity is less than the
velocity of light, may describe idealized thin walls encountered
in astronomical observations \cite{cmb}.
 If $\omega^{2}>{\bf q}^{2}$ then we show that the plane waves are
Lorentz transformations of homogeneous solutions of Einstein
equations with an ideal fluid as a source (Lorentz boosts to an
arbitrary frame of the well-known homogeneous solutions obtained
in the frame moving with the fluid, see \cite{peebles} for an
application of such boosts). If $\omega^{2}<{\bf q}^{2}$ then the
phase velocity is less than the velocity of light. The plane waves
can be considered as Lorentz transformations of static solutions
of Einstein equations (with a static fluid) . Another explicit
travelling wave solution results when the source is a free  scalar
field or a scalar field with an exponential interaction.

We discuss in more detail the case $\omega^{2}={\bf q }^{2}$. In
this case the travelling plane wave moves with the velocity of
light. We show that a massless free field or an electromagnetic
field of a plane wave can be a source of the gravitational
travelling  plane wave. As another source we  consider a particle
 with a space-time dependent mass. We believe that the
travelling gravitational waves moving with the velocity of light
may be relevant near the strong sources where the linear
approximation to Einstein equations is not sufficient ( the
well-known plane fronted exact gravitational waves are solutions
of sourceless Einstein equations \cite{solutions} ). In general,
it is not simple (see \cite{flanagan}) to divide the metric
resulting from various sources into the radiative and
non-radiative parts. Nevertheless, both parts influence the
geodesic motion of a test body, hence are measurable. The
conformally flat metrics play a distinguished role because of
their Lorentz covariance \cite{infeld}. In such a metric the
electromagnetic and gravitational radiation always propagates with
the velocity of light. For this reason the conformally flat
metrics are an interesting object of investigation.

 The plan of the paper is
the following. In sec.2 we discuss the conformal flat metrics. In
sec.3 we show that if the source is an ideal fluid then there are
solutions of Einstein equations in the form of a travelling plane
wave. In sec.4 we discuss scalar fields as a source of the
energy-momentum tensor. We show that  in the case of the
travelling wave there is a (dispersion) relation between $\omega$
and $\vert{\bf q}\vert$. In sec.5 we solve the geodesic equation
in a gravitational field of the travelling plane-wave. In sec.6 we
discuss Lagrangian models of the energy-momentum leading to plane
wave solutions with $\omega=\vert{\bf q}\vert$. In sec.7 we obtain
some explicit formulas for  the metric of the plane wave with
$\omega=\vert{\bf q}\vert$. In sec.8 we summarize the results.
\section{The  conformally flat metric}

We consider the conformally flat metric in four space-time
dimensions (we set the velocity of light $c=1$)
\begin{equation}
ds^{2}=a(x)^{2}(d\tau^{2}-d{\bf
x}^{2})=a(x)^{2}\eta^{\mu\nu}dx_{\mu}dx_{\nu},
\end{equation}where $\eta $ is the Minkowski metric.
 Then, the components of the Einstein tensor $G_{\mu\nu}=R_{\mu\nu}-\frac{1}{2}g_{\mu\nu}R$ are
 \cite{blaschke}\cite{solutions}
\begin{equation}
G_{00}=3(a^{-1}\partial_{\tau}a)^{2}+(a^{-1}\nabla
a)^{2}-2a^{-1}\triangle a,
\end{equation}
\begin{equation}
G_{0j}=4a^{-2}\partial_{\tau}a\partial_{j}a-2a^{-1}\partial_{\tau}\partial_{j}a,
\end{equation}
\begin{equation}
G_{jk}=4a^{-2}\partial_{j}a\partial_{k}a+\delta_{jk}a^{-2}((\partial_{\tau}a)^{2}-(\nabla
a
)^{2})-2a^{-1}(\partial_{j}\partial_{k}a+\delta_{jk}(\partial_{\tau}^{2}-\nabla^{2})a).
\end{equation}
We assume that the fields as well as the scale factor depend only
on the plane wave phase \begin{equation} \xi=\omega\tau -{\bf
qx}=\eta^{\mu\nu}q_{\mu}x_{\nu},
\end{equation} where $(q_{\mu})=(\omega,{\bf q})$.
Then (where $q^{2}={\bf q}^{2}$)
\begin{equation}
G_{00}=(3\omega^{2}+q^{2})a^{-2}(\frac{da}{d\xi})^{2}-2q^{2}a^{-1}\frac{d^{2}a}{d\xi^{2}}
 \end{equation}
The remaining components of the Einstein tensor as functions of
$\xi$ are
\begin{equation}
G_{0j}=\omega
q_{j}a^{-2}\Big(4(\frac{da}{d\xi})^{2}-2a\frac{d^{2}a}{d\xi^{2}}\Big),
\end{equation}
\begin{equation}
G_{jk}=a^{-2}\Big(4q_{j}q_{k}+\delta_{jk}(\omega^{2}-q^{2})\Big)(\frac{da}{d\xi})^{2}
-2a^{-1}\Big(q_{j}q_{k}+\delta_{jk}(\omega^{2}-q^{2})\Big)\frac{d^{2}a}{d\xi^{2}}.
 \end{equation}

If $\partial_{\mu}a$ is not a null vector  then Einstein equations
can be treated as an identity defining an energy-momentum tensor
for a relativistic viscous fluid \cite{eckart}\cite{brevik}(this
way Einstein equations are treated in the quark-gluon plasma
\cite{fluids}). We write $T_{\mu\nu}=(8\pi G)^{-1}G_{\mu\nu}$
(where $G$ is the Newton constant) then $G_{\mu\nu}$ can be
expressed as
\begin{equation}
(8\pi
G)^{-1}G_{\mu\nu}=(\rho+p)u_{\mu}u_{\nu}-pg_{\mu\nu}+\Pi_{\mu\nu},
\end{equation}
where the fluid velocity is defined by
\begin{equation}
u_{\mu}=\partial_{\mu}a
\Big(\partial_{\mu}a\partial^{\mu}a\Big)^{-\frac{1}{2}}.
\end{equation}
The energy density is
\begin{equation}
8\pi
G\rho=3a^{-2}g^{\mu\nu}\partial_{\mu}a\partial_{\nu}a-g^{\mu\nu}\Pi_{\mu\nu}
\end{equation}
the pressure
\begin{equation}
8\pi G
p=a^{-2}g^{\mu\nu}\partial_{\mu}a\partial_{\nu}a+g^{\mu\nu}\Pi_{\mu\nu},
\end{equation}
where
\begin{equation}
8\pi G\Pi_{\mu\nu}=-2a^{-1}\partial_{\mu}\partial_{\nu}a.
\end{equation}
The conservation law $(G^{\mu\nu})_{;\mu}=0$ can be decomposed
into a continuity equation for the fluid
$u_{\mu}(T^{\mu\nu})_{;\nu}=0$ and  a Navier-Stokes type equation
$(g_{\sigma\mu}-u_{\sigma}u_{\mu})(T^{\mu\nu})_{;\nu}=0$.

For space-like $q^{\mu}$ the square root in eq.(10) makes no
sense. Then, we define
\begin{equation}
u_{\mu}=\partial_{\mu}a
\Big(-\partial_{\mu}a\partial^{\mu}a\Big)^{-\frac{1}{2}}.
\end{equation}
\begin{equation}
8\pi
G\rho=-3a^{-2}g^{\mu\nu}\partial_{\mu}a\partial_{\nu}a+g^{\mu\nu}\Pi_{\mu\nu}
\end{equation}

\begin{equation}
8\pi
Gp=a^{-2}g^{\mu\nu}\partial_{\mu}a\partial_{\nu}a+g^{\mu\nu}\Pi_{\mu\nu},
\end{equation}Hence
\begin{equation}
(8\pi
G)^{-1}G_{\mu\nu}=(\rho-p)u_{\mu}u_{\nu}-pg_{\mu\nu}+\Pi_{\mu\nu},
\end{equation}
It follows from eq.(9) that for general $a$ the Einstein tensor
$G_{\mu\nu}$ has the form of the energy-momentum tensor of a
viscous fluid. However, if $a$ depends only on $\xi$ then we can
write $G_{\mu\nu}$ as the energy-momentum of an ideal fluid
\begin{equation}
(8\pi
G)^{-1}G_{\mu\nu}=(\tilde{\rho}+\tilde{p})\tilde{u}_{\mu}\tilde{u}_{\nu}-\tilde{p}g_{\mu\nu},
\end{equation}
where
\begin{equation}\tilde{u}_{\mu}=q_{\mu}a(\omega^{2}-q^{2})^{-\frac{1}{2}},
\end{equation}
\begin{equation}
8\pi
G\tilde{p}=(a^{-4}(\frac{da}{d\xi})^{2}-2a^{-3}\frac{d^{2}a}
{d\xi^{2}})
(\omega^{2}-q^{2}),
\end{equation}
\begin{equation}
8\pi G\tilde{\rho}=3a^{-4}(\frac{da}{d\xi})^{2}(\omega^{2}-q^{2}).
\end{equation}
From eqs.(20)-(21) we can obtain the continuity equation
conveniently expressed in terms of the e-fold time $\nu=\ln(a)$.
Then
\begin{equation}
\partial_{\nu}\tilde{\rho}+3(\tilde{\rho}+\tilde{p})=0
\end{equation}
We have two equations (20)-(21) for one function $a$.If we define
$\rho(a)$ in eq.(21) then we obtain $a(\xi)$ and inserting it in
eq.(20) we determine $p(a)$ . It follows from eq.(22) that an
equation of state $p=f(\rho)$ determines $\rho(a)$ and $p(a)$
which are consistent with eqs.(20)-(21). With  $p(a)$ and
$\rho(a)$ satisfying eqs. (20)-(21) Einstein equations with
$T_{\mu\nu}$ defined on the rhs of eq.(18)(with $u_{\mu}$ of
eq.(19)) will be satisfied.

Let us still mention a modification of eqs.(18)-(21) if
$q>\omega$. From eqs.(14)-(17)
\begin{equation}
(8\pi
G)^{-1}G_{\mu\nu}=(\tilde{\rho}-\tilde{p})\tilde{u}_{\mu}\tilde{u}_{\nu}-\tilde{p}g_{\mu\nu},
\end{equation}
where
\begin{equation}\tilde{u}_{\mu}=q_{\mu}a(q^{2}-\omega^{2})^{-\frac{1}{2}},
\end{equation}
\begin{equation}
8\pi
G\tilde{p}=(3a^{-4}(\frac{da}{d\xi})^{2}+2a^{-3}\frac{d^{2}a}
{d\xi^{2}})
(q^{2}-\omega^{2}),
\end{equation}
\begin{equation}
8\pi G\tilde{\rho}=(-a^{-4}(\frac{da}{d\xi})^{2}+2a^{-3}\frac{d^{2}a}
{d\xi^{2}})
(q^{2}-\omega^{2}).
\end{equation}

 The energy-momentum $T_{\mu\nu}$ of the gravitational field (proportional to
 $G_{\mu\nu}$)
is covariantly conserved. We introduce the energy-momentum of the
matter field $Q_{\mu\nu}$ so that
\begin{equation}
8\pi G t_{\mu\nu}=G_{\mu\nu}+Q_{\mu\nu}
\end{equation}
satisfies
\begin{equation}
\partial_{\mu}t^{\mu\nu}=0.
\end{equation}
We obtain
\begin{equation}
8\pi Gt_{\mu\nu}=
(a^{-2}(\frac{da}{d\xi})^{2}-2a^{-1}\frac{d^{2}a}{d\xi^{2}})
(q_{\mu}q_{\nu}-\eta_{\mu\nu}(\omega^{2}-q^{2}))
\end{equation} and

\begin{equation}
8\pi G Q_{\mu\nu}=3a^{-2}(\frac{da}{d\xi})^{2}q_{\mu}q_{\nu}.
\end{equation}
If $(q^{\mu})=(1,{\bf 0})$ then $8\pi GQ_{00}=3
(a^{-1}\partial_{\tau}a)^{2}$ as in the model with the
energy-momentum $Q_{00}$ of a homogeneous fluid.

\section{Ideal fluids}
We have ten Einstein equations
\begin{equation} G_{\mu\nu}=R_{\mu\nu}-\frac{1}{2}g_{\mu\nu}R=8\pi
G T_{\mu\nu}
\end{equation}for a single conformal factor $a$. Eqs.(31) are not of the same type.
$G_{jk}$ equations are hyperbolic whereas equations for $G_{00}$
and $G_{0j}$ are parabolic (first order in time derivatives). For
$G_{jk}$ we can insert arbitrary initial conditions for time
derivatives of $a$ whereas the initial value for the time
derivative of $a$ in the equations for $G_{00}$ and $G_{0j}$ is
determined by the initial conditions for $T_{\mu\nu}$. The
distinction between the $G_{\mu\nu}$ equations is not explicit for
the $\xi$ equations (which are of the second order in $\xi$).
However, if the $G_{00}$ equation is to have the $q=0$ limit then
we have to choose a proper initial condition for
$\partial_{\xi}a$. If we solve one of Eqs.(31) then the remaining
equations determine the other components of the energy-momentum
tensor. It remains an open problem whether the energy-momentum
tensor defined this way follows from a Lagrangian field theory. As
an example let us consider the $\delta_{jk}$ part of $G_{jk}$
equation

\begin{equation}
8\pi
G\tilde{p}=(a^{-4}(\frac{da}{d\xi})^{2}-2a^{-3}\frac{d^{2}\xi}{d\xi^{2}})(\omega^{2}-q^{2})
\end{equation}
Choosing
\begin{equation}
\tilde{p}=p_{0}a^{-3-3w}(\omega^{2}-q^{2})
\end{equation}
we can integrate eq.(32) with the result
\begin{equation}
a^{-2}(\frac{da}{d\xi})^{2}=(k_{0}-8\pi
G(3w)^{-1}p_{0})a^{-1}+8\pi Gp_{0}(3w)^{-1}a^{-1-3w}
\end{equation} where $k_{0}$ is an arbitrary constant. By differentiation of eq.(34) we obtain
\begin{equation}
2a^{-1}\frac{d^{2}a}{d\xi^{2}}=(k_{0}-8\pi G(3w)^{-1})a^{-1}+8\pi
Gp_{0}(3w)^{-1}(1-3w)a^{-1-3w}
\end{equation}
We can then calculate all  terms $G_{\mu\nu}$ and $T_{\mu\nu}$ as
functions of $a$. In particular, from eq.(21)
\begin{equation}
\tilde{\rho}=3(\omega^{2}-q^{2})(8\pi G)^{-1}(k_{0}-8\pi
G(3w)^{-1})a^{-3}+p_{0}w^{-1}a^{-3-3w}
\end{equation}
As an equation of state we obtain
\begin{equation}
\tilde{\rho}(\tilde{p})=b\tilde{p}^{-\frac{1}{1+w}}+w^{-1}\tilde{p}
\end{equation}where
\begin{displaymath}b=
3(\omega^{2}-q^{2})(8\pi G)^{-1}(k_{0}-8\pi
G(3w)^{-1})p_{0}^{-\frac{1}{1+w}}. \end{displaymath}We have the
linear relation
\begin{equation}
\tilde{p}=w\tilde{\rho} \end{equation} if and only if
\begin{equation} k_{0}=8\pi G(3w)^{-1}
\end{equation}
The result (34) shows that a power-law  for $\tilde{p}$ admits
only the equation of state (37) as a solution of eq.(22) (with
eq.(38) as a particular case).

If we solve eq.(21) with the Ansatz \begin{equation}
\rho=\rho_{0}a^{-3-3w}
\end{equation}
 and insert the solution in eq.(20)  then we obtain
the result (38) and the initial condition (39). On the other hand
if we assume eq.(38) then from eq.(22) we obtain the power-law
behaviour (33) and (40). Then, the solution of eq.(31) is a
standard extension $\tau\rightarrow \xi $ of the homogeneous
solution. We obtain this way  a Lorentz transformation of the
homogeneous solution if $\tau$ and $\xi$ are related by a Lorentz
transformation (see the end of the next section).
\section{Scalar field as a source}
Let us consider the  Lagrangian for scalar fields
\begin{equation}
L=\frac{1}{2}g^{\mu\nu}\partial_{\mu}\phi\partial_{\nu}\phi-V
\end{equation}
The energy-momentum tensor is
\begin{equation}
T_{\mu\nu}=\partial_{\mu}\phi\partial_{\nu}\phi-g_{\mu\nu}L
\end{equation}
We investigate  a soluble model of the power-law inflation
\cite{exp}
\begin{equation}
V=\lambda\exp(\alpha\phi)+V_{0},
\end{equation}where $V_{0}$ is a constant.
We assume that the fields depend only on $\xi$. Then, the Lagrange
equation for $\phi$ reads
\begin{equation}
(\omega^{2}-q^{2})\frac{d}{d\xi}a^{2}\frac{d}{d\xi}\phi=-a^{4}V^{\prime}
\end{equation}
We solve the $G_{0j}$ Einstein equations first with (from eq.(42))
\begin{equation}
T_{0j}=\omega q_{j}(\frac{d\phi}{d\xi})^{2}
\end{equation}
assuming
\begin{equation}
(\frac{d\phi}{d\xi})^{2}=\kappa^{2}a^{-r},
\end{equation}where $\kappa$ is a constant.
Let us note that if $V=V_{0}$ ($\lambda=0$ in eq.(43) ) then
$r=4$. Eq.(31) for $G_{0j}$ can be integrated (with the Ansatz
(46); we choose the initial condition $a(0)=1$)
\begin{equation}
a^{-2}(\frac{da}{d\xi})^{2}=(k_{0}-8\pi
G\kappa^{2}(r+2)^{-1})a^{2} +8\pi G\kappa^{2}(r+2)^{-1}a^{-r}
\end{equation}with an integration constant $k_{0}$.
It follows from eq.(47) that\begin{equation}
2a^{-1}\frac{d^{2}a}{d\xi^{2}}=4(k_{0}-8\pi
G\kappa^{2}(r+2)^{-1})a^{2} +8\pi
G\kappa^{2}(2-r)(r+2)^{-1}a^{-r}.
\end{equation}
We can insert the results (47)-(48) into the remaining Einstein
equations and into the Lagrange equations for $\phi$. If
$\lambda=0$ then the remaining Einstein equations have the
solution if
\begin{equation}
3(k_{0}-8\pi G\kappa^{2}(r+2)^{-1})(\omega^{2}-q^{2})=8\pi GV_{0},
\end{equation}
whereas the Lagrange equations (44) require $r=4$ in eq.(46) (
solutions $a(\xi)$ of eq.(47) are discussed in sec.7 below
eq.(93)) .

When $\lambda\neq 0 $ then we must have $V_{0}=0$ (hence there is
zero on the rhs of eq.(49)). Then, the Lagrange equations (44) and
the remaining Einstein equations are solved if
\begin{equation}
a=\sigma\exp(\beta\phi),
\end{equation}\begin{equation}
(r+2)\beta^{2}=8\pi G,
\end{equation}
\begin{equation}
\frac{\alpha}{\beta}=-r-2.
\end{equation}
\begin{equation}
\kappa^{2}(\omega^{2}-q^{2})=2\lambda\frac{2+r}{4-r}\end{equation}
Then, in addition  the dispersion relation
\begin{equation}
\omega^{2}=q^{2}+\frac{2\lambda}{\kappa^{2}}\sigma^{r+2}\frac{2+r}{4-r}
\end{equation}
must be satisfied. We obtain
\begin{equation}
a^{\frac{r}{2}}=\frac{1}{2}r\kappa\beta\xi
\end{equation}
$G_{\mu\nu}$ is covariant with respect to Lorentz transformations
$L$. Hence, if the energy-momentum tensor on the rhs of eq.(31) is
also Lorentz covariant then \begin{equation}
(L^{-1})_{\mu}^{\alpha}
(L^{-1})_{\nu}^{\beta}G_{\alpha\beta}(Lx)=8\pi
G(L^{-1})_{\mu}^{\alpha}
(L^{-1})_{\nu}^{\beta}T_{\alpha\beta}(Lx).
\end{equation}
Note that $\xi=\eta^{\mu\nu}q_{\mu}x_{\nu}$ is Lorentz invariant.
In a special Lorentz frame it may take the form
$\xi=\omega_{0}\tau$ where
$\omega_{0}^{2}=\eta^{\mu\nu}q_{\mu}q_{\nu}$ if $q_{\mu}$ is
time-like or $\xi=-q_{10}x^{1}$ where
$q_{10}^{2}=-\eta^{\mu\nu}q_{\mu}q_{\nu}$ if $q_{\mu}$ is
space-like. We may solve Einstein equations (31)in a special frame
(with a special choice of $q$). Then, on the basis of eq.(56) they
will hold true for a general $\xi$.

As an example we apply the Lorentz transformation when $V=0$ and
$\omega^{2}\neq q^{2}$ in eq.(44). Then\begin{equation}
\partial_{\xi}\phi=\kappa a^{-2}
\end{equation} with a certain constant $\kappa$. Choose $\omega=1$
and ${\bf q}=0$. We solve eq.(6) for $G_{00}$
($a\simeq\tau^{\frac{1}{2}}$). In the resulting solution $a(\tau)$
we replace $\tau$ by $\xi=\gamma \tau-v\gamma x_{1}$ where
$(q_{\mu})=(\gamma,v\gamma,0,0)$ with
$\gamma=(1-v^{2})^{-\frac{1}{2}}$ coming from the Lorentz
transformation (an application of such a boost of a homogeneous
solution is discussed in \cite{peebles}). We check that
\begin{equation}
a^{2}=\sqrt{\frac{16\pi G}{3}}\kappa\xi
\end{equation}
solves all of 10 Einstein equations (31) ( $T_{\mu\nu} $ of
eq.(42) with $V=0$ ) transformed to an arbitrary frame by eq.(56).

In the case with an exponential interaction if $r<4$  and
$\lambda>0$ we have from eq.(54) $\omega>q$. We could solve the
homogeneous Einstein equations ($q=0$, as in \cite{exp}). Then, by
a boost we obtain the inhomogeneous solution (55) in the form of
the plane wave with the phase velocity $\frac{\omega}{q}>1$ larger
than the velocity of light. If $r>4$ (large $\alpha$) and
$\lambda>0$ then $\omega<q$. We can look for static one
dimensional solutions of Einstein equations with only $x^{1}\neq
0$.  Then, by a boost we obtain a general plane wave solution (55)
with the phase velocity $\frac{\omega}{q}<1$ (less than the
velocity of light). This is the standard way how solitons are
derived from static solutions in field theory of a scalar field
\cite{solitons}. It is well-understood that if the plane wave
moves with the velocity ${\bf u}$ less than the velocity of light
then in the frame moving with the velocity ${\bf u}$ the wave
looks static.

\section{Test-body motion } We consider a
geodesic motion of a body under the influence of the conformally
flat gravity. The geodesic equation is
\begin{equation}
\frac{d^{2}x^{\mu}}{ds^{2}}+\Gamma^{\mu}_{\alpha\nu}\frac{dx^{\alpha}}{ds}\frac{dx^{\nu}}{ds}=0,
\end{equation}
where
\begin{displaymath}
ds^{2}=g_{\alpha\beta}dx^{\alpha}dx^{\beta}=a^{2}d\tau^{2}(1-(\frac{d{\bf
x}}{d\tau})^{2}).
\end{displaymath}
Inserting the Christoffel symbols $\Gamma$ we derive a simple
equation for $\xi$
\begin{equation}
\frac{d^{2}\xi}{ds^{2}}+2(\frac{d\xi}{ds})^{2}\frac{d}{d\xi}\ln
a=-\frac{1}{2}(\omega^{2}-q^{2})\frac{d}{d\xi}a^{-2}.\end{equation}
Denoting $\nu=\frac{d\xi}{ds}$ we can integrate eq.(60) with the
result
\begin{equation}
\nu^{2}=(K_{0}-(\omega^{2}-q^{2})a_{0}^{2})a^{-4}+(\omega^{2}-q^{2})a^{-2}
\end{equation} where $K_{0}$ and $a_{0}$ are constants of integration. From eq.(61) we can calculate $a(s)$
\begin{equation}
\int da(\frac{da}{d\xi})^{-1}
\Big((K_{0}-(\omega^{2}-q^{2})a_{0}^{2})a^{-4}+(\omega^{2}-q^{2})a^{-2}\Big)^{-\frac{1}{2}}=s,
\end{equation}
where $\frac{da}{d\xi}$ is determined by eq.(34) or
(47).Subequently, we can obtain $\xi (s)$ when $a(\xi)$ is known .

The velocity $\frac{d\xi}{ds}$ is expressed by the phase velocity
${\bf u}$ ( for the velocity $\vert{\bf u}\vert <1$)
\begin{displaymath}
a\nu =a\frac{d\xi}{ds}=(\omega-{\bf qu})\frac{1}{\sqrt{1-{\bf
u}^{2}}}
\end{displaymath}
with
\begin{displaymath}
{\bf u}=\frac{d{\bf x}}{d\tau}.
\end{displaymath}
The phase velocity is related to the coordinate velocity in proper
time
\begin{displaymath}
\frac{d{\bf x}}{ds}=a^{-1}{\bf u}\frac{1}{\sqrt{1-{\bf u}^{2}}}
\end{displaymath}
In order to determine the motion beyond the $\xi$ plane let us
define
\begin{displaymath}
v^{\mu}=\frac{dx^{\mu}}{ds}-q^{\mu}(\omega^{2}-q^{2})^{-1}\frac{d\xi}{ds}
\end{displaymath}
Inserting $v^{\mu}$ into the geodesic equations (59) we obtain
\begin{displaymath}
\frac{dv^{\mu}}{ds}+2v^{\mu}\frac{d\ln(a)}{ds}=0
\end{displaymath}
Hence,
\begin{displaymath}
v^{\mu}(s)=v^{\mu}(0)a(\xi(s))^{-2}a(\xi(0))^{2}
\end{displaymath}
where $\xi(s)$ is determined from the solution of eq.(60). In
principle, we could detect the source of gravity observing its
action upon a test body. However, it would be difficult to
separate the astrophysical sources from the local ones.
\section{Einstein
equations for $\omega=q$} When  $\omega=q$  then the travelling
plane waves generated by the energy-momentum $T_{\mu\nu}$ move
with the velocity of light. We consider various sources of such
waves: massless scalar fields, electromagnetic fields and a
relativistic particle with a continuous spectrum of mass. These
fields produce a deformation of the space-time which propagates
with the velocity of light. The waves will interact with massive
test-bodies and charged particles. They carry an energy and
momentum which are conserved because if $\omega=q$ then in
eqs.(6)-(8) $\partial_{\mu}G^{\mu\nu}=0$ hence also
$\partial_{\mu}T^{\mu\nu}=0$.

 As a
source of the energy-momentum in this section we consider a scalar
field with the Lagrangian
\begin{equation}
L_{h}(\phi)=h(W),
\end{equation}
where $h$ is an arbitrary function and
\begin{displaymath}
W=\frac{1}{2}\partial_{\mu}\phi\partial^{\mu} \phi
\end{displaymath}
The electromagnetic field with the Lagrangian
\begin{equation}
L_{em}=-\frac{1}{4}F_{\mu\nu}F^{\mu\nu}
\end{equation}
 provides a source moving with a velocity of light. We discuss
also  a relativistic particle with $x$-dependent energy density
$m$
 with the
Lagrangian
\begin{equation}
L_{p}(x)=\frac{1}{2}\int ds
g_{\mu\nu}(X)\frac{dX^{\mu}}{ds}\frac{dX^{\nu}}{ds}m(x)\delta(X(s)-x).
\end{equation}
 From  the Lagrangian $-\frac{R}{8\pi
G} +L$ we obtain Einstein equations (31).
 The scalar field Lagrangian  defines the energy-momentum tensor

\begin{equation} T_{h}^{\mu\nu}
=h^{\prime}\partial^{\mu}\phi\partial^{\nu}\phi-g^{\mu\nu}h
\end{equation}From the Lagrangian (64) we obtain

\begin{equation}
T_{\mu\nu}=g^{\alpha\beta}F_{\mu\alpha}F_{\nu\beta}-\frac{1}{4}g_{\mu\nu}F^{\alpha\beta}F_{\alpha\beta}.
\end{equation}The energy-momentum of the relativistic particle (65) is defined as
 \begin{equation} T_{\mu\nu}=\int
ds\frac{dX_{\mu}}{ds}\frac{dX_{\nu}}{ds}\delta(X(s)-x)m(x).
\end{equation}

 The Lagrangian equations for the scalar field
(under the assumption that the fields depend only on $\xi$) are
\begin{equation}
\partial_{\mu}(h^{\prime}\sqrt{-g}g^{\mu\nu}\partial_{\nu}\phi)=
(\omega^{2}-q^{2})\partial_{\xi}(h^{\prime}a^{2}\partial_{\xi}\phi)=0.
\end{equation}
It follows from eq.(69) that if $\omega^{2}=q^{2}$ then for every
function $h$ any $\phi(\xi)$ satisfies the scalar wave equation.
Note that if $\omega^{2}=q^{2}$ then $W=0$. Hence, $(h(W)=h(0)$
and  $h^{\prime}(W)=h^{\prime}(0)$  are constants in the
energy-momentum (66) . The electromagnetic field satisfies the
Maxwell equations
\begin{equation}
(F^{\mu\nu})_{;\mu}=0
\end{equation} (together with
$(\epsilon^{\mu\nu\alpha\beta}F_{\alpha\beta})_{;\mu}=0$). The
Lagrange equations for the particle read
\begin{equation}\begin{array}{l}
\frac{d}{ds}\Big(a^{-2}(x)\frac{dX_{\mu}}{ds}\delta(X(s)-x)m(x)\Big)=
\cr
a(x)^{-2}\eta^{\alpha\beta}\frac{dX_{\alpha}}{ds}\frac{dX_{\beta}}{ds}m(x)
\frac{\partial}{\partial
X^{\mu}(s)}\delta(X(s)-x)\end{array}\end{equation}where $x=\xi$.
If  $q_{0}=\vert {\bf q}\vert$ then eq.(71) and the constraint
$x=\xi$ are solved by
 \begin{equation}
 X_{\mu}(s)=q_{\mu} s+x.
 \end{equation} Then,
 $\eta^{\mu\nu}X_{\mu}q_{\nu}=\eta^{\mu\nu}x_{\mu}q_{\nu}=\xi$
. Under the assumption that $a$ depends only on $\xi$ (5) with
$\omega^{2}=q^{2}$ eqs.(2)-(4) can be expressed in the form
\begin{equation}
G_{\mu\nu}=q_{\mu}q_{\nu}
\Big(4a^{-2}(\frac{da}{d\xi})^{2}-2a^{-1}\frac{d^{2}a}{d\xi^{2}}\Big),
\end{equation}
 where $(q_{\mu})=(q,{\bf q})$.

The components of the scalar field energy-momentum tensor (on
solutions (69)) are
\begin{equation}
T_{00}=h^{\prime}\omega^{2}(\frac{d\phi}{d\xi})^{2}-a^{2}h,
\end{equation}\begin{equation}
T_{0j}=h^{\prime}q_{j}\omega(\frac{d\phi}{d\xi})^{2},
\end{equation}\begin{equation}
T_{jk}=h^{\prime}q_{j}q_{k}(\frac{d\phi}{d\xi})^{2}+a^{2}h\delta_{jk}.
\end{equation} Comparing eqs.(74)-(76) with
eq.(73) we can see that if 00 component of Einstein equations is
satisfied then the remaining components will be satisfied if
\begin{equation}
h(0)=0.
\end{equation}
We assume the normalization $h^{\prime}(0)=1$ as for the free
massless scalar field (when $h(W)=W$).

After an insertion of the solution (72) the  particle
energy-momentum (68) is
\begin{equation}
T_{\mu}^{\nu}=q_{\mu}q^{\nu}q^{-1} m(\xi).
\end{equation}
In the case of the electromagnetic field (64) it is known (see
,e.g.,\cite{LL} ) that for the plane wave solutions
\begin{equation} T_{\mu}^{\nu}(\xi)=\omega^{-2}\rho_{em}(\xi)q_{\mu}q^{\nu},
\end{equation}where $\rho_{em}$ is the electromagnetic energy
density. Summarizing, it follows from eqs.(73) and (74)-(79) that
in the Lagrangian models of this section it is sufficient to solve
the 00 component of Einstein equations. After the solution of the
$00$ component the remaining equations will be satisfied.

We have no a priori restriction on the density
$\rho(\xi)q_{\mu}q_{\nu}=a^{-2}T_{\mu\nu}$ in the formulas
(66)-(71)
 for $T_{\mu\nu}$. $T_{\mu\nu}(\xi)$  can  be expressed as a
function of $a$. For a comparison with standard models we choose
this $a$-dependence in a power-law form as it is obtained for
ideal fluids with the equation of state $p=w\rho$ (where $p$ is
the pressure and  $w$ is a constant). It is sufficient to restrict
ourselves to one of the models (74)-(79). In the scalar field
model  the 00 component of Einstein equations reads
\begin{equation}
4a^{-2}(\frac{da}{d\xi})^{2}-2a^{-1}\frac{d^{2}a}{d\xi^{2}}=8\pi
Gq^{-2} T_{00}=8\pi G(\frac{d\phi}{d\xi})^{2}
\end{equation}
where we express $\frac{d\phi}{d\xi}$ as a function of $a$.

 Eq.(80) is solved with
the initial conditions $a(\xi=0)=a_{0}$ and
$\frac{da}{d\xi}(\xi=0)=u_{0}$.
 Let
$K=u^{2}$ then eq.(80)  can be expressed as
\begin{equation}
\frac{dK}{da}-4a^{-1}K=-8\pi Gq^{-2}a T_{00}.
\end{equation}
Eq.(81) can be integrated
\begin{equation}
(\frac{da}{d\xi})^{2}=K=a^{4}k_{0}-8\pi
Gq^{-2}a^{4}\int_{a_{0}}^{a}b^{-3} T_{00}(b)db,
\end{equation}where we introduced an integration constant $k_{0}\geq 0$  related to
$u_{0}=\pm a_{0}^{2}\sqrt{k_{0}}$.

If $T_{00}=0$ then $a=a_{0}=const$ is a solution of eq.(82) with
$k_{0}=u_{0}=0$. However, there is also a non-trivial solution. It
can be checked that
\begin{equation}
\frac{1}{a}=\frac{1}{a_{0}}\pm\sqrt{k_{0}}\xi
\end{equation}
is a solution of the Einstein equation (31) with the
energy-momentum tensor $T_{\mu\nu}=0$. This is a special case of
the plane fronted gravitational waves (flat polarization) when
they are conformally flat (as noted in
\cite{maartens}\cite{tooper}).

 We assume that $T_{00}$ is a power of $a$
\begin{equation}
\rho=a^{-2}T_{00}=a^{-2}\kappa^{2}q^{2}(\frac{d\phi}{d\xi})^{2}=
 \rho_{0}a^{-3-3w}.
\end{equation}
We may  assume the $a$-dependence because a function of $\xi$ can
be expressed as a function of $a$ (if  $a(\xi)$ is invertible).
So, we may assume that $\phi(\xi)=\tilde{\phi}(a)$ (we skip
"tilde" further on). Then
\begin{equation}
(\frac{d\phi}{d\xi})^{2}=(\frac{d\phi}{da})^{2}(\frac{da}{d\xi})^{2}.
\end{equation}
Hence,
\begin{equation} (\frac{d\phi}{da})^{2}=
\kappa^{-2}\rho_{0}a^{-1-3w}K^{-1},
\end{equation}
where from eqs.(80)-(82)
\begin{equation}
K(a)=(k_{0}-2\alpha^{2}(3w+3)^{-1})a^{4}+2\alpha^{2}(3w+3)^{-1}a^{1-3w},
\end{equation}where
\begin{equation}
\alpha^{2}= 4\pi G\rho_{0}\kappa^{-2}
\end{equation}
and we set $a_{0}=1$  ($a_{0}>0$ has no physical meaning it just
rescales coordinates). Taking the square root in eq.(87) and
integrating we can calculate $\phi$ as a function of $a$.

Note that if we know $\phi(a)$ then eq.(85) can be expressed in
another form suitable for integration
\begin{equation}
(\frac{da}{d\xi})^{2}=K_{0}a^{4}\exp\Big(-8\pi G\int
da(\frac{d\phi}{da})^{2} a\Big).
\end{equation}
 Eq.(82) can be expressed as
\begin{equation} (\frac{da}{d\xi})^{2}=-V(a),
\end{equation} where $V=-K$ is  a potential
of a particle moving with the kinetic energy $K$ on the half-line
$a\geq 0$ with the total energy $E=0$ . The motion is possible in
the range of $a$ such that $-V\geq 0$. The evolution $a(\xi)$
based on eqs.(82) and (90) is discussed in the next section.

\section{Elementary solutions of Einstein equations} Taking the
square root of eq.(87) and integrating we obtain the equation for
$a$
\begin{equation}
\int_{1}^{a}db K(b)^{-\frac{1}{2}}=\pm \xi.
\end{equation}
We consider the cases when the integral (91) can be expressed by
elementary functions or elliptic functions (such integrals have
been discussed in \cite{coq}\cite{dabrowski}\cite{szydlo}).

Let $w=1$ in eq.(84) (stiff matter \cite{stiff}). Denote
\begin{displaymath} 16\sigma_{s}=k_{0}-\frac{1}{3}\alpha^{2}
\end{displaymath}
and consider  $A=a^{2}$. Then
\begin{equation}
K=a^{-2}(16\sigma_{s}a^{6}+\frac{1}{3}\alpha^{2}).
\end{equation}
If $\sigma_{s}> 0$
\begin{equation}\int_{1}^{A}
dA\Big(4A^{3}+\frac{\alpha^{2}}{48\sigma_{s}}\Big)^{-\frac{1}{2}}=8\sqrt{\sigma_{s}}(\xi+\xi_{0})
\end{equation}
then the integral is expressed by the Weierstrass elliptic
function ${\cal
P}(8\sqrt{\sigma_{s}}(\xi+\xi_{0}),0,-\frac{\alpha^{2}}{48\sigma_{s}})$
\cite{coq}\cite{dabrowski}. We can extend the upper limit in
eq.(93) to infinity showing that if $\sigma_{s}>0$ then $a$
achieves infinity for a finite $\xi$ (then the energy density
tends to zero). If $\sigma _{s}<0$  then the range of $a$ is
bounded by the requirement $K\geq 0$.

 For $w=\frac{1}{3}$ (relativistic matter)
\begin{equation}
K=(k_{0}-\frac{1}{2}\alpha^{2})a^{4}+\frac{1}{2}\alpha^{2}.
\end{equation}
Hence, again the integral (91) is expressed by an elliptic
function. If $k_{0}-\frac{1}{2}\alpha^{2}>0$ then there is an
explosion at finite $\xi$, whereas if
$k_{0}-\frac{1}{2}\alpha^{2}<0$ then $a$ varies in a bounded
interval.

Let us consider now negative $w$. First, $w=-\frac{1}{3}$ (cosmic
string or coasting cosmology \cite{coasting})
\begin{equation}
K(a)=a^{4}(k_{0}-\alpha^{2})+ \alpha^{2}a^{2}.
\end{equation}
Let
\begin{equation}
4\sigma=\alpha^{2}-k_{0}.
\end{equation}
Then, the integral (91) gives
\begin{equation}
a=\alpha^{2}\exp(-\alpha(\xi+\xi_{0}))\Big(\alpha^{2}\sigma+\exp(-2\alpha(\xi+\xi_{0}))\Big)^{-1},
\end{equation}
where $\xi_{0}$ is chosen is such a way as to satisfy the initial
condition $a(0)=1$. If $\sigma<0$ then $a\rightarrow \infty $ at
finite $\xi$ (then the energy density tends to zero ) . If
$\sigma>0$ then $a$ is a bounded function ($a\rightarrow 0$ when
$\xi\rightarrow \pm\infty$).

If $w=-\frac{2}{3}$ (domain wall) then
\begin{equation}
K=(k_{0}-2\alpha^{2})a^{4}+2\alpha^{2}a^{3}.
\end{equation}
Eq.(91) gives
\begin{equation}
a=2\alpha^{2}\Big(\alpha^{4}(\xi+\xi_{0})^{2}-k_{0}+2\alpha^{2}\Big)^{-1}.
\end{equation}
If $2\alpha^{2}<k_{0}$ then $a$ explodes at finite $\xi$ (the
energy density  tends to 0).If $2\alpha^{2}>k_{0}$ then $a$ is a
bounded function.

Let $w=-\frac{4}{3}$ (phantom matter)
\begin{equation}
K=(k_{0}+2\alpha^{2})a^{4}-2\alpha^{2}a^{5}
\end{equation}
From the integral (91) we obtain
\begin{equation}\begin{array}{l}
\xi+\xi_{0}=-(k_{0}+2\alpha^{2})^{-1}a^{-1}\sqrt{k_{0}+2\alpha^{2}-2\alpha^{2}a}\cr
+\alpha^{2}(k_{0}+2\alpha^{2})^{-\frac{3}{2}}
\ln\Big(\Big(\sqrt{k_{0}+2\alpha^{2}}-\sqrt{k_{0}+2\alpha^{2}-2\alpha^{2}a}\Big)
\cr\Big(\sqrt{k_{0}+2\alpha^{2}}+\sqrt{k_{0}+2\alpha^{2}-2\alpha^{2}a}\Big)^{-1}\Big)
\end{array}\end{equation}
$\xi_{0}$ is chosen to satisfy the initial condition $a(\xi=0)=1$.
$a$ varies in the interval $[1,1+\frac{k_{0}}{2\alpha^{2}}]$.

If $w=-\frac{5}{3}$ then
\begin{equation}
K=(k_{0}+\alpha^{2})a^{4}-\alpha^{2}a^{6}.
\end{equation}
The integral is again an elementary function. This problem is
similar to the previous one so we do not write down (rather
complicated ) explicit formula. If $w=-2$ then $a$ varies in a
bounded interval, it can  be expressed by an elliptic function. If
$w<-2$ then the integral (91) can be discussed by means of the
methods of ref.\cite{szyd} which allow to detect some elementary
solutions for integer values of $w$.

\section{Summary}
We have discussed some exact inhomogeneous solutions  of Einstein
equations in the form of conformally flat travelling plane waves.
Such gravitational deformations of space-time can propagate with
the velocity smaller
 than the velocity of light. They can disturb measurements of some quantities with a
global meaning such as the Hubble constant.We can test the
presence of inhomogeneities  on the basis of a geodesic motion of
a test body . If the source of the disturbance  is an ideal fluid
with time-like velocity then the plane waves can be considered as
a transformation of the homogeneous expanding solution to the
moving frame. If the fluid has space-like velocity (a static
source) then the plane wave has the phase velocity smaller than
the velocity of light. In the frame moving with the wave its
amplitude and energy is static. It can describe a measurable wave
propagation. In the case when the phase velocity is equal to the
velocity of light the travelling waves may contain gravitational
waves produced by a massless scalar field
 or an electromagnetic field. The metric is not
asymptotically flat. Hence, the gravitational field is not of the
type of pure radiation (it may contain a non-radiation
background). The electromagnetic field which is produced when
 neutron stars merge can be a source of plane
conformally flat gravitational  waves. These waves could possibly
be detected in a different experimental set-up than the one so far
prepared for a detection of the transverse (TT) gravitational
waves. From the mathematical point of view the conformally flat
solutions are distinguished by their covariance with respect to
the Lorentz transformations. Using this covariance we can obtain
non-homogeneous solutions from homogenous solutions of Einstein
equations (with Lorentz covariant sources) or time-dependent
solutions from static ones.

\end{document}